\documentclass[runningheads]{llncs}
\usepackage{cite}
\usepackage{amsmath,amssymb,amsfonts}
\usepackage{algorithmic}
\usepackage{graphicx}
\usepackage{textcomp}
\usepackage{xcolor}
\usepackage{multirow}

\usepackage{float}
\usepackage{hyperref}
\usepackage{algorithm}
\usepackage{lipsum}
\usepackage[utf8]{inputenc}

\usepackage[caption=false, labelformat=simple, font=footnotesize]{subfig}

\usepackage{array}
\newcolumntype{P}[1]{>{\centering\arraybackslash}p{#1}}

\begin{document}
\newcommand{\name}{SiGAT}
\newcommand{\V}{\mathcal{V}}
\newcommand{\E}{\mathcal{E}}
\newcommand{\G}{\mathcal{G}}
\newcommand{\nedge}{k}
\newcommand{\mb}{\mathbf}
\newcommand{\eg}{e.g., }
\newcommand{\ie}{i.e., }

\title{Signed Graph Attention Networks}
%
%
\author{Junjie Huang\inst{1,2,*} \and
Huawei Shen\inst{1} \and
Liang Hou\inst{1,2} \and Xueqi Cheng\inst{1}}
\authorrunning{	J.Huang et al.}
%
\institute{CAS Key Laboratory of Network Data Science and Technology, \\Institute of Computing Technology, Chinese Academy of Sciences, Beijing, China\and
University of Chinese Academy of Sciences, Beijing, China\\
\email{\{huangjunjie17s, shenhuawei, houliang17z, cxq\}@ict.ac.cn}}
%
\maketitle              
\begin{abstract}
Graph or network data is ubiquitous in the real world, including social networks, information networks, traffic networks, biological networks and various technical networks. The non-Euclidean nature of graph data poses the challenge for modeling and analyzing graph data. Recently, Graph Neural Network (GNNs) are proposed as a general and powerful framework to handle tasks on graph data, e.g., node embedding, link prediction and node classification. As a representative implementation of GNNs, Graph Attention Networks (GATs) are successfully applied in a variety of tasks on real datasets. However, GAT is designed to networks with only positive links and fails to handle signed networks which contain both positive and negative links. In this paper, we propose Signed Graph Attention Networks (SiGATs), generalizing GAT to signed networks. SiGAT incorporates graph motifs into GAT to capture two well-known theories in signed network research, i.e., balance theory and status theory. In SiGAT, motifs offer us the flexible structural pattern to aggregate and propagate messages on the signed network to generate node embeddings. We evaluate the proposed SiGAT method by applying it to the signed link prediction task. Experimental results on three real datasets demonstrate that SiGAT outperforms feature-based method, network embedding method and state-of-the-art GNN-based methods like signed graph convolutional network (SGCN).

\end{abstract}

\section{Introduction}\label{sec:introduction}

Graph Neural Networks (GNNs) have been receiving more and more research attentions, achieving good results in many machine learning tasks \eg semi-supervised node classification\cite{kipf2016semi}, network embedding\cite{kipf2016variational} and link prediction.  
GNNs introduce neural networks into graph data by defining convolution\cite{kipf2016semi}, attention\cite{velickovic2017graph} and other mechanisms. 
However, previous GNNs methods mainly focus on undirected and unsigned networks (\ie networks consisting of only positive edges). 
How to apply GNNs to signed, directed, weighted and other complex networks is an important research direction. 

With the growing popularity of online social media, many interactions are generated by people on the Web. 
Most of these interactions are positive relationships, such as friendship, following, support and approval. 
Meanwhile, there are also some negative links that indicate disapproval, disagreement or distrust.
Some recent researches about signed social networks suggest that negative links bring more valuable information over positive links in various tasks\cite{leskovec2010predicting}. 
In the early years, the methods for modeling signed networks were mainly based on feature extraction\cite{leskovec2010predicting}, matrix factorization\cite{hsieh2012low} and network embedding\cite{wang2017signed,kim2018side, sodhani2019attending}. They have achieved good performances in tasks such as signed link prediction and node classification. 
The core of modeling signed networks is two important sociological theories, \ie balance theory and status theory. 
Derr et al.\cite{derr2018signed} proposed a signed graph convolution network (SGCN) method to model signed networks. 
It utilizes balance theory to aggregate and propagate the information. 
However, it only applies to the undirected signed networks.


In this paper, we try to model the directed signed network into graph attention network (GAT).
To the best of our knowledge, it's the first time to introduce GAT to directed signed networks. Attention mechanism can characterize the different effects of different nodes on the target node, \eg A is is more kind to me than B.

The main contributions of this paper are as follows:

\begin{itemize}
    \item We analyze the key elements of the signed link prediction problem. Triads motifs are used to describe two important sociological theories, \ie balance theory and status theory.
    \item We introduce the GAT to model the signed network and design a new motif-based GNN model for signed networks named SiGAT.
    \item We conduct experiments on some real signed social network data sets to demonstrate the effectiveness of our proposed framework SiGAT.
\end{itemize}

The rest of paper is organized as follows. In section~\ref{sec:related_work}, we review some related work.
We propose our model SiGAT in section~\ref{sec:methods}, which consists of introducing related social theory, incorporating these theories into motifs and finally presenting our Signed Graph Attention Network (SiGAT).
In section~\ref{sec:experiments}, experiments are performed to empirically evaluate the effectiveness of our framework for learning node embedding.
We discuss the results of some signed network embedding and SGCN, analyze the impact of hyper-parameters in our model.
Finally, we conclude and discuss future work in section~\ref{sec:conclusion}.

\section{Related work}\label{sec:related_work}

\subsection{Signed Network Embedding}
Signed social networks are such social networks in which social ties can have two signs: positive and negative\cite{leskovec2010signed}. 
To mine signed networks, many algorithms have been developed for tasks such as community detection, node classification, link prediction and spectral graph analysis.
In particular, \cite{leskovec2010signed} proposed a feature engineering-based methods to do the signed link prediction tasks. 
Recently, with the development of network representation learning\cite{perozzi2014deepwalk, grover2016node2vec, tang2015line}, researchers begin to learn low-dimensional vector representations for a social network.
For signed network embedding methods, SNE\cite{yuan2017sne} adopts the log-bilinear model and incorporates two signed-type vectors to capture the positive or negative relationship of each edge along the path. 
SiNE\cite{wang2017signed} designs a new objective function guided by social theories to learn signed network embedding, it proposed to add a virtual node to enhance the training process. 
SIDE\cite{kim2018side} provides a linearly scalable method that leverages balance theory along with random walks to obtain the low-dimensional vector for the directed signed network. 
SIGNet\cite{islam2018signet} combines balance theory with a specialized random and new sampling techniques in directed signed networks. 
These methods are devoted to defining an objective function that incorporates sociological theory and then using some machine learning techniques \eg sampling and random walks to optimize look-up embedding. 

\subsection{Graph Neural Networks}
GNNs have achieved a lot of success in the semi-supervised node classification task\cite{kipf2016semi}, graph classification\cite{gilmer2017neural} and other graph analysis problems. 
The concept of graph neural network (GNN) was first proposed in \cite{scarselli2009graph}, which extended neural networks for processing the data represented in graph domains. 
Recently, researchers tried to apply convolution\cite{kipf2016semi}, attention\cite{velickovic2017graph} , auto-encoder\cite{kipf2016variational} and other mechanisms into graphs.

GCN in\cite{kipf2016semi} is designed with the focus to learning representation at the node level in the semi-supervised node classification task. 
GraphSAGE\cite{hamilton2017inductive} extends it to the large inductive graphs.
GAE\cite{kipf2016variational} firstly integrates GCN into a graph auto-encoder framework. 
This approach is aimed at representing network vertices into a low-dimensional vector space by using neural network architectures in unsupervised learning tasks.
GAT\cite{velickovic2017graph} applies attention mechanisms to GNNs. 
Compared to the convolution, the attention mechanism can describe the influence of different nodes on the target node. 
On the semi-supervised task of the node, it shows better results than GCN.
SGCN\cite{arinik2017signed} introduced GNNs to a signed network for the first time. 
It designed a new information aggregation and propagation mechanism for the undirected signed network according to the balance theory. 
However, the majority of current GNNs tackle with static homogeneous graphs\cite{wu2019comprehensive}, how to apply GNNs into other complex networks(\eg directed signed networks) is still a challenge.

\section{Methods}\label{sec:methods}
\subsection{Signed Network Theory}

Balance theory and status theory play an important role in analyzing and modeling signed networks. 
These theories can enable us to characterize the differences between the observed and predicted configurations of positive and negative links in online social networks\cite{leskovec2010predicting}. 
\begin{figure}[!ht]
    \vspace{-20px}
    \subfloat[Illustration of structural balance theory]{%
      \centering
        \includegraphics[width=0.50\textwidth]{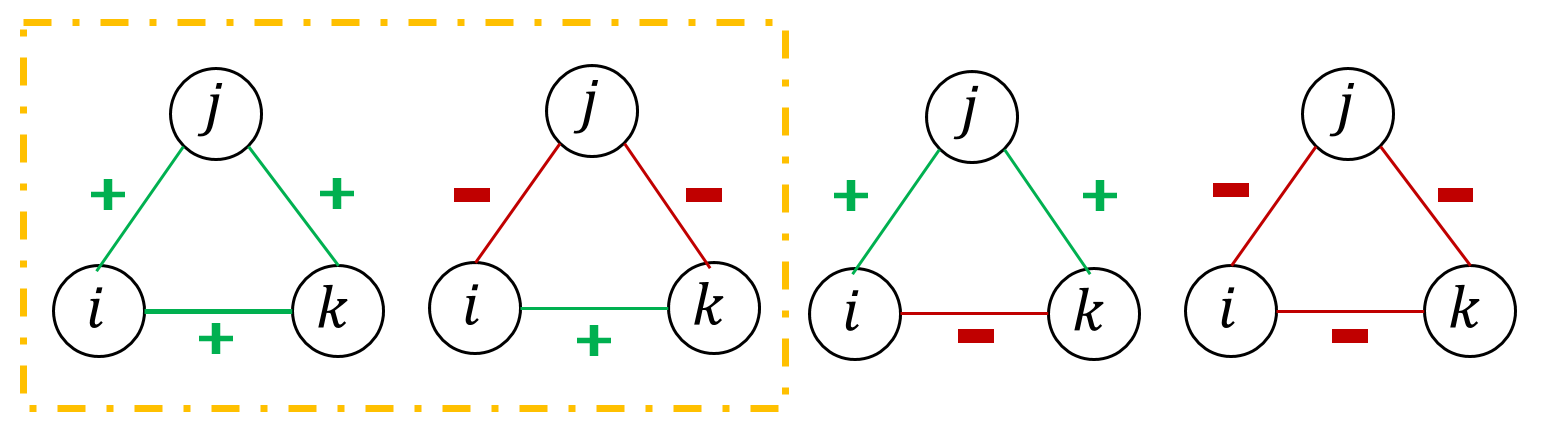}
        \label{fig:balance_theory}
    }
    \label{fig:1a}
    \subfloat[Illustration of status theory]{%
         \centering
        \includegraphics[width=0.50\textwidth]{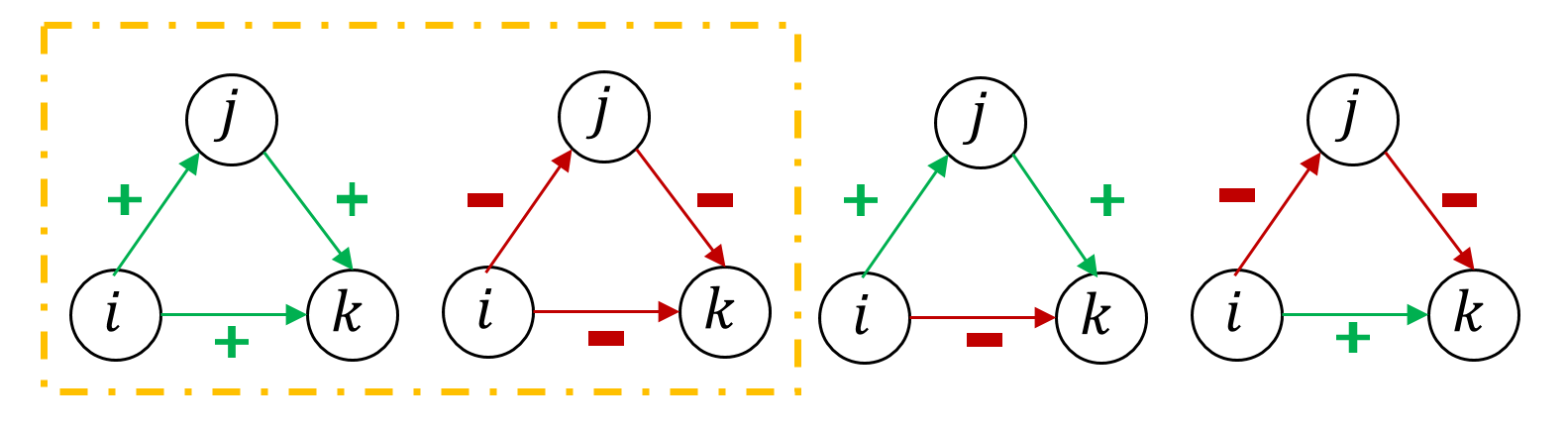}
        \label{fig:status_theory}   
    }
    \label{fig:1b}
    \caption{The balance theory and status theory}
    \vspace{-10px}
\end{figure}

\subsubsection{Balance Theory}

Balance theory originated in social psychology in the mid-20th-century\cite{heider1946attitudes}.
It was initially intended as a model for undirected signed networks. 
All signed triads with an even number of negative edges are defined as balanced. 
In other words, for the four triangles in Fig.\ref{fig:balance_theory}, the triangles which all three of these users are friends or only one pair of them are friends are defined as balanced \ie the first two triads are balanced.
Balance theory suggests that people in a social network tend to form into a balanced network structure. 
It can be described as ``the friend of my friend is my friend" and ``the enemy of my enemy is my friend".
This theory has a wide range of applications in the field of sociology. \cite{leskovec2010predicting} found that balance theory has widely existed in social networks. 
It was proven that if a connected network is balanced, it can be split into two opposing groups and vice versa\cite{easley2010networks}. 

\subsubsection{Status Theory}

Status theory is another key social psychological theory for signed network.
It is based on directed signed networks.
It supposes directed relationship labeled by a positive sign ``+" or negative sign ``-"  means target node has a higher or lower status than source node\cite{tang2012inferring}. 
For example, a positive link from A to B, it does not only mean ``B is my friend" but also ``I think B has a higher status than I do.".
For the triangles in Fig.\ref{fig:status_theory}, the first two triangles satisfy the status ordering and the latter two do not satisfy it. 
For the first triangle, when $\mathrm{Status(j)} > \mathrm{Status(i)}$ and $\mathrm{Status(k)} > \mathrm{Status(j)}$, we have $\mathrm{Status(k)} > \mathrm{Status(i)}$. 
Based on status theory, it can be found that ``opinion leaders" have higher social status (manager or advisor) than ordinary users\cite{tang2012inferring}. 


Based on the analysis of the above two theories on social network datasets, \cite{leskovec2010predicting} extracts a total of 23 features of two given nodes $u$ and $v$ and then uses logistic regression to train and infer the positive and negative of edges. 
Experiments show that these features can help to achieve good results in machine learning tasks. 
These two theories can be described in triangles. These directed edges, signed edges and triangles can all be defined as motifs of a network\cite{milo2002network}. 
In signed social networks, nodes not only spread the message/information/influence by just directed positive or negative relationship but also triad motifs. 


\subsection{Signed Motifs}

Since positive neighbors and negative neighbors should have different effects on the target node, they should obviously be treated separately.
Furthermore, for directed signed networks, the direction also contains some knowledge, \eg status theory.
Regard as previous discussions about social-psychological theories, we think that triads also have different effects. 
As we said, these different effects can be described by different motifs.
Therefore, we propose a unified framework to effectively model signed directed networks based on motifs.

\begin{figure}[H]
    \centering
    \includegraphics[width=0.45\textwidth]{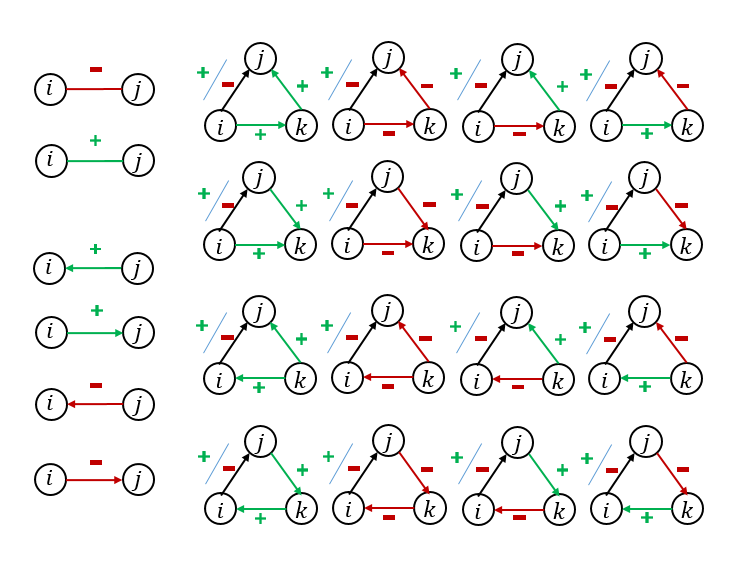}
    \caption{38 different motifs in SiGAT, mean different influences from node $j$ to node $i$} 
    \label{fig:motifs}
    \vspace{-10px}
\end{figure}

In our framework, we define positive/negative neighbors (2 motifs), positive /negative with direction neighbors (4 motifs) and 32 different triangle motifs in Fig.\ref{fig:motifs}. 
We first extract these motifs and then apply them to our model. The method of effectively extracting these motifs can be found in \cite{schank2005finding}.


\subsection{Signed Graph Attention Networks}
GAT\cite{velickovic2017graph} introduces attention mechanism into the graph.
It uses a weight matrix to characterize the different effects of different nodes on the target node.
GAT firstly computes the $\alpha_{ij}$ for node $i$ and node $j$ by the attention mechanism $\vec{\bf a}$ and $\text{LeakyReLU}$ nonlinearity (with negative input slope $\alpha = 0.2$) as:
\begin{equation}
    \label{eq:gat1}
    \alpha_{ij} = \frac{\exp\left(\text{LeakyReLU}\left(\vec{\bf a}^T[{\bf W}\vec{h}_i\|{\bf W}\vec{h}_j]\right)\right)}{\sum_{k\in\mathcal{N}_i} \exp\left(\text{LeakyReLU}\left(\vec{\bf a}^T[{\bf W}\vec{h}_i\|{\bf W}\vec{h}_k]\right)\right)},
\end{equation}
where $\cdot^T$ represents transposition and $\|$ is the concatenation operation, $\mathcal{N}_i$ are the neighborhoods of node $i$, $\bf W$ is the weight matrix parameter, $\vec{h}_i$ is the node feature of node $i$.

After computing the normalized attention coefficients, a linear combination of the features is used and served to the final output features for every node:
\begin{equation}\label{eqnatt}
	\vec{h}'_i = \sigma\left(\sum_{j\in\mathcal{N}_i} \alpha_{ij} {\bf W}\vec{h}_j\right).
\end{equation}

\begin{figure*}[!ht]
    \centering
    \includegraphics[width=0.85\textwidth]{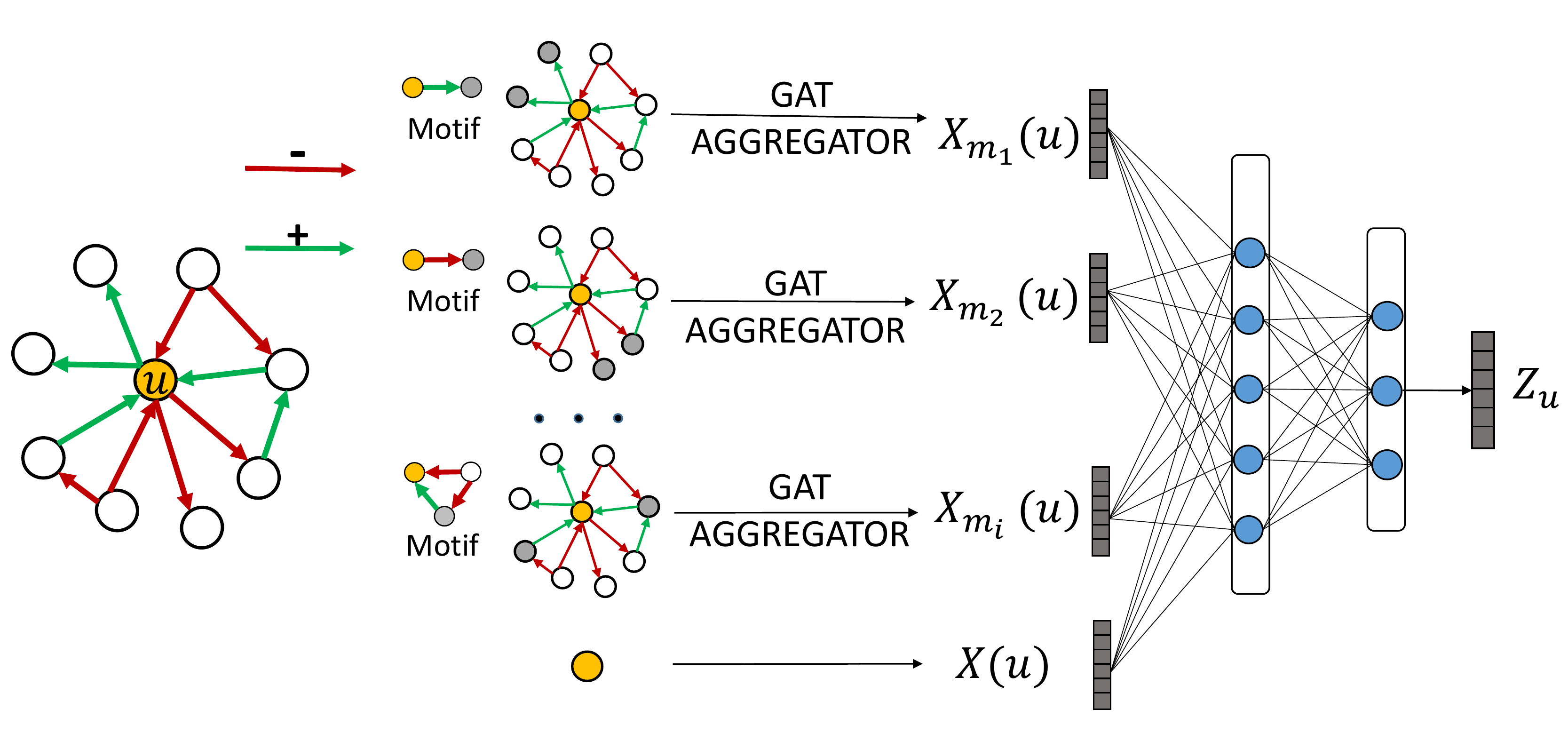}
    \caption{For a signed network, the target $u$ is the orange node. Red/Green links mean positive/negative links, the arrows indicate the directions.
    We define different motifs and obtain the corresponding neighborhoods in them. Under a motif definition, the gray color nodes are the neighborhoods, we use GAT to describe the influences from these neighborhoods. After aggregating the information from different motifs, we use these $\mb{X}_{m_i}(u)$ and $\mb{X}(u)$ to get the representation $\mb{Z}_u$.  }
    \label{fig:sigat}

\end{figure*}

Based on previous discussions, we proposed our SiGAT model in Fig.\ref{fig:sigat}. For a node $u$, in different motifs, we get different neighborhoods. For example, in $v\rightarrow^{+} u$ motifs definition, the gray color nodes $v$ are the neighborhoods, we use GAT to describe the influence from these neighborhoods. After aggregating the information from different motifs, we concatenate these $\mb{X}_{m_i}(u)$ and $\mb{X}(u)$ to a two-layer fully connected neural network to generate the representation $\mb{Z}_u$ for node $u$, $m_i$ is the $i$-th motif.
We also take the scalability into considerations and propose a mini-batch algorithm like GraphSAGE\cite{hamilton2017inductive}. 

The details are described in Algorithm~\ref{alg:algorithm1}. 
First, for each node $u$, we initialize embeddings $\mb{X}(u)$ as features. 
We extracted new graphs under different motif definitions using the motif graph extract function $F_{m_i}$, based on motif list which is given before.  
In other word, for each target node $u$, we get its neighbor nodes $\mathcal{N}_{m_i}(u)$ under motif $m_i$ definition. 
Then, for every motif $m_i$, we use a GAT model $\textsc{GAT-AGGREGATOR}_{m_i}$ with the parameter ${\bf W}_{m_i},{\bf \vec{a}}_{m_i}$ as follows to get the message $\mb{X}_{m_i}(u)$ in this motif:
\begin{equation}
    \label{eq:gat1}
    \alpha_{uv}^{m_i} = \frac{\exp\left(\text{LeakyReLU}\left(\vec{\bf a}_{m_i}^T[{\bf W}_{m_i}{\mb{X}(u)}\|{\bf W}_{m_i}{\mb{X}(v)}]\right)\right)}{\sum_{k\in\mathcal{N}_{m_i}(u)} \exp\left(\text{LeakyReLU}\left(\vec{\bf a}_{m_i}^T[{\bf W}_{m_i}{\mb{X}(u)}\|{\bf W}_{m_i}{\mb{X}(k)}]\right)\right)},
\end{equation}
\begin{equation}\label{eqnatt}
	\mb{X}_{m_i}(u) = \sum_{v\in\mathcal{N}_{m_i}(u)} \alpha_{uv}^{m_i} {\bf W}_{m_i}\mb{X}(v).
\end{equation}

\begin{algorithm}[!ht]
    \caption{SiGAT embedding generation (forward) algorithm}
    \label{alg:algorithm1}
    \begin{algorithmic}[1]
        \renewcommand{\algorithmicrequire}{\textbf{Input:}}
        \renewcommand{\algorithmicensure}{\textbf{Output:}}
        \REQUIRE {
                Sigend directed Graph $G(V,E, s)$;\\
                Motifs list $\mathcal{M}$; Epochs $T$; Batch Size $B$; \\
                Motifs graph extract function $F_{m_i}, \forall m_i \in \mathcal{M}$;\\
                Aggregator $\textsc{GAT-AGGREGATOR}_{m_i}$ with the parameter ${\bf W}_{m_i},{\bf \vec{a}}_{m_i}$, $\forall m_i \in \mathcal{M}$;\\
                Weight matrices $\mathbf{W_1}, \mathbf{W_2}$ and bias $\mathbf{b_1}, \mathbf{b_2}$; \\
                Non-linearity function $\mathrm{Tanh}$
                };\\

        \ENSURE  Node representation $\mb{Z}_{u}, \forall u \in V$
        \\ 
        \STATE{$\mb{X}(u)\leftarrow \text{random}(0, 1), \forall u \in V$}
        \STATE{$G_{m_i}\leftarrow F_{m_i}(G), \forall m_i \in \mathcal{M}$}
        \STATE{$\mathcal{N}_{m_i}(u) \leftarrow \{v| (u,v) \in G_{m_i} \}, \forall m_i \in \mathcal{M}, \forall u \in V$}
        \FOR{$epoch=1,...,T$}
            \FOR{$batch=1,...,|V|/B$}
                 \STATE{$\mathcal{B}\leftarrow V_{(batch-1)\times B+1:batch\times B}$}
                
                \FOR{$u \in \mathcal{B}$}
                    \FOR{$m_i \in \mathcal{M}$}
                        \STATE{
                        $\mb{X}_{m_i}(u) \leftarrow \textsc{GAT-AGGREGATOR}_{m_i}(\{\mb{X}_v, \forall v \in \mathcal{N}_{m_i}(v)\})$\;
                        }
                    \ENDFOR
                    \STATE{
                    $\mb{X}'(u)\leftarrow  \textsc{CONCAT} ( \mb{X}(u), \mb{X}_{m_1}(u),..., \mb{X}_{m_{|\mathcal{M}|}}(u) )$
                    }
                    \STATE{
                    $\mb{Z}_{u} \leftarrow  \mathbf{W_2} \cdot \mathrm{Tanh} (\mathbf{W_1} \cdot \mb{X}'(u) + \mathbf{b_1} ) + \mathbf{b_2}$
                    }

                \ENDFOR
            \ENDFOR

            \ENDFOR
        \RETURN {$\mb{Z}_{u}$}
    \end{algorithmic}
\end{algorithm}

Finally, we concatenate all the messages from different motifs with $\mb{X}(u)$ to a two-layer fully connected neural network to obtain the final embedding $\mb{Z}_u$ for the target node $u$.
Here, $\mathbf{W_1}, \mathbf{W_2}, \mathbf{b_1}, \mathbf{b_2}$ are the parameters for the 2-layer neural network.

After computing the loss, we propagate the gradient back and update the parameters.
In our models, we define an unsupervised loss function in Eq.\ref{eq:loss-function} which reflects friend embeddings are similar, and the enemy embeddings are dissimilar. 
\begin{equation}\label{eq:loss-function}
    J_{\G}(\mb{Z}_u) = -\sum_{v^+\in \mathcal{N}(u)^+}\log\left(\sigma(\mb{Z}^\top_u\mb{Z}_{v^+}) \right) - Q\sum_{v^-\in \mathcal{N}(u)^-}\log\left(\sigma(-\mb{Z}^\top_u\mb{Z}_{v^-})\right),
\end{equation}
where $\sigma$ is the sigmoid function, $\mathcal{N}(u)^+$ is the set of positive neighborhoods of node $u$, $\mathcal{N}(u)^-$ is the set of negative neighborhoods of node $u$ and $Q$ is the balanced parameter for the unbalanced positive and negative neighborhoods.

\section{Experiments}\label{sec:experiments}
\subsection{Link Sign Prediction and Dataset Description}
Link sign prediction is the task of predicting unobserved signs of existing edges in the test dataset given train dataset. It's the most fundamental signed network analysis task\cite{leskovec2010predicting}.

We do experiments on three real-world signed social network datasets. \eg Bitcoin-Alpha\footnote{http://snap.stanford.edu/data/soc-sign-bitcoin-alpha.html}, Slashdot\footnote{http://snap.stanford.edu/data/soc-sign-Slashdot090221.html} and Epinions\footnote{http://snap.stanford.edu/data/soc-sign-epinions.html}. 

Bitcoin-Alpha\cite{kumar2016edge} is from a website where users can trade with Bitcoins. Because the Bitcoin accounts are anonymous, the users need to build online trust networks for their safety. Members of Bitcoin-Alpha rate other members in a scale of -10 (total distrust) to +10 (total trust) in steps of 1, which can help to prevent transactions with fraudulent and risky users. In this paper, we treat the scores greater than 0 as positive and others as negative. 
Slashdot\cite{leskovec2010signed} is a technology-related news website where users are allowed to tag each other as friends or foes. In other words, it is a common signed social network with friends and enemies labels.
Epinions\cite{leskovec2010signed} was a product review site. Members of the site can indicate their trust or distrust of the reviews of others. The network reflects people's opinions to others.
In these networks, the edges are inherently directed\cite{leskovec2010signed}.

The statistics of three datasets are summarized in Table~\ref{tb:dataset}. We can see that positive and negative links are imbalanced in these datasets.
\begin{table}[!ht]
\vspace{-10px}
    \centering
    \caption{Statistics of Three Datasets}
    \label{tb:dataset}
         \scalebox{0.8}{
        \begin{tabular}{P{3cm}P{2cm}P{2cm}P{2cm}P{2cm}P{2cm}} 
        \hline
        Dataset  & \# nodes & \# pos links & \# neg links & \% pos ratio  \\ 
        \hline
        Bitcoin-Alpha & 3,783 & 22,650 & 1,536 & 93.65 \\
        Slashdot & 82,140   & 425,072      & 124,130      & 77.40  \\
        Epinions & 131,828   & 717, 667  &123,705      & 85.30    \\
        \hline
    \end{tabular}
    }
   
\vspace{-20px}
\end{table}
\subsection{Baselines}

To validate the effectiveness of SiGAT, we compare it with some state-of-the-art baselines and one simple version SiGAT$_{\mathrm{ +/-}}$. The baselines are as follows:

\begin{itemize}
    \item Random: It generates $d$ dimensional random values, $Z_j = (z_1, z_2, ...,z_{d}), z_i \in [0.0, 1.0)$. It can be used to show the logistic regression's ability in this task.
    
    \item Unsigned network embedding: We use some classical unsigned network embedding methods (\eg DeepWalk\cite{perozzi2014deepwalk},  Node2vec\cite{grover2016node2vec}, LINE\cite{tang2015line}) to validate the importance of signed edges.
    
    \item Signed network embedding: We use some signed network embedding methods(\eg SIDE\cite{kim2018side}, SiNE\cite{wang2017signed}, SIGNet\cite{islam2018signet}) to show the effectiveness of modeling signed edges.

    \item SGCN\cite{derr2018signed}: It makes a dedicated and principled effort that utilizes balance theory to correctly aggregate and propagate the information across layers of a signed GCN model. It is the latest attempt to introduce GNNs to signed networks.

    \item FeExtra\cite{leskovec2010predicting}: This method extracts two parts, a total of 23 features from the signed social network. 
    
    \item SiGAT$_{\mathrm{+/-}}$: We made some simplifications, considering only the positive and negative neighbors, the model can verify if the attention mechanism works.
\end{itemize}

For a fair comparison, the embedding dimension $d$ of all methods is set to 20 which is closed to FeExtra\cite{leskovec2010predicting}.
We use the authors released code for DeepWalk, Node2vec, LINE, SIDE, SiNE and SIGNet. For SGCN, we use the code from github\footnote{https://github.com/benedekrozemberczki/SGCN}. 
We follow the authors recommended settings in Slashdot and Epinions and try to fine-tune parameters to achieve the best results in Bitcoin-Alpha.
Like previous works\cite{kim2018side, wang2017signed, derr2018signed}, we first use these methods to get node representations. 
For edge $e_{ij}$, we concatenate these two learned representation $z_i$ and $z_j$ to compose an edge representation $z_{ij}$. 
After that, we train a logistic regression classifier on the training set and use it to predict the edge sign in the test set. 
We perform 5-fold cross-validation experiments to get the average scores as the same as \cite{kim2018side}. 
Each training set is used to train both embedding vectors and logistic regression model.
Our models\footnote{https://github.com/huangjunjie95/SiGAT} were implemented by PyTorch with the Adam optimizer ($\mathrm{Learning~Rate}=0.0005$, $\mathrm{Weight~Decay} = 0.0001$, $\mathrm{Batch~Size} = 500$). 
The epoch was selected from $\{10, 20, 50, 100\}$ by 5-fold cross-validation on the training folds.

\subsection{Results}

\begin{table}[!ht]
    \centering
    \caption{The results of Signed Link Prediction on three datasets}
    \label{tb:experiment-result}
\setlength{\leftskip}{-40pt}
    \scalebox{0.8}{
  \begin{tabular}{c|c|c|ccc|ccc|c|ccc}
\hline
\multicolumn{2}{c|}{} & \multicolumn{1}{c|}{Random} & \multicolumn{3}{c|}{Unsigned Network Embedding} & \multicolumn{3}{c|}{Signed Network Embedding} & \multicolumn{1}{c|}{\begin{tabular}[c]{@{}c@{}}Feature\\ Engineering\end{tabular}} & \multicolumn{3}{c}{Graph Neural Network} \\ \hline
Dataset                        & Metric   & Random & Deepwalk & Node2vec & LINE   & SiNE   & SIDE   & SIGNet & FExtra  &SGCN & SiGAT$_{\mathrm{+/-}}$ &SiGAT  \\ \hline
\multirow{4}{*}{Bitcoin-Alpha} & Accuracy & 0.9365 & 0.9365 & 0.9274 & 0.9350 & 0.9424 & 0.9369 & 0.9443 & 0.9477 & 0.9351 & 0.9427 & \textbf{0.9480} \\
                              & F1 & 0.9672 & 0.9672 & 0.9623 & 0.9662 & 0.9699 & 0.9673 & 0.9706 & 0.9725 & 0.9658 & 0.9700 & \textbf{0.9727} \\
                              & Macro-F1 & 0.4836 & 0.4836 & 0.5004 & 0.5431 & 0.6683 & 0.5432 & 0.7099 & 0.7069 & 0.6689 & 0.6570 & \textbf{0.7138} \\
                              & AUC & 0.6395 & 0.6435 & 0.7666 & 0.7878 & 0.8788 & 0.7832 & \textbf{0.8972} & 0.8887 & 0.8530 & 0.8699 & 0.8942 \\ \hline
\multirow{4}{*}{Slashdot}      & Accuracy & 0.7740 & 0.7740 & 0.7664 & 0.7638 & 0.8269 & 0.7776 & 0.8391 & 0.8457 & 0.8200 & 0.8331 & \textbf{0.8482} \\
                              & F1 & 0.8726 & 0.8726 & 0.8590 & 0.8655 & 0.8921 & 0.8702 & 0.8984 & \textbf{0.9061} & 0.8860 & 0.8959 & 0.9047 \\
                              & Macro-F1 & 0.4363 & 0.4363 & 0.5887 & 0.4463 & 0.7277 & 0.5469 & 0.7559 & 0.7371 & 0.7294 &  0.7380 &\textbf{0.7660} \\
                              & AUC & 0.5415 & 0.5467 & 0.7622 & 0.5343 & 0.8423 & 0.7627  & 0.8749 & 0.8859 & 0.8440 & 0.8639 &\textbf{0.8864}\\ \hline
\multirow{4}{*}{Epinions}      & Accuracy & 0.8530 & 0.8518 & 0.8600 & 0.8262 & 0.9131 & 0.9186 & 0.9116 & 0.9206 & 0.9092 & 0.9124 & \textbf{0.9293} \\
                              & F1 & 0.9207 & 0.9198 & 0.9212 & 0.9040 & 0.9502 & 0.9533 & 0.9491 & 0.9551 & 0.9472 & 0.9498 & \textbf{0.9593} \\
                              & Macro-F1 & 0.4603 & 0.4714 & 0.6476 & 0.4897 & 0.8054 & 0.8184 & 0.8065 & 0.8075 & 0.8102 & 0.8020 & \textbf{0.8449} \\
                              & AUC & 0.5569 & 0.6232 & 0.8033 & 0.5540 & 0.8882 & 0.8893 & 0.9091 & \textbf{0.9421} & 0.8818 & 0.9079 & 0.9333 \\ \hline
\end{tabular}
}
\end{table}

We report the average Accuracy, F1, macro-F1 and AUC in Table~\ref{tb:experiment-result}. 
We have bolded the highest value of each row. From Table~\ref{tb:experiment-result}, we can find that:
\begin{itemize}
    \item For signed networks, itself is the positive and negative imbalance.
    Given random embedding, logistic regression can learn a part of the linear features to a certain extent.
    \item After using unsigned network embedding methods, all metrics have been increased. Node2Vec has made an obvious improvement, and DeepWalk and LINE have different performances on different datasets.  
    \item SiNE, SIDE, and SIGNet are designed for the signed network; the result is significantly higher than other unsigned network embedding methods. These algorithms modeled some related sociological theories and had achieved good results for signed network analysis. SIGNet showed more stable and better results in our experiments. Besides, The performance of SIDE is not as good as that in \cite{kim2018side}. This can be due to the different pre-processing processes and low dimension parameter. 
    \item  Although FeExtra was proposed earlier, it has performed well because of its successful use of relevant sociological theory. However, it should be pointed out that this method relies on feature engineering to extract features manually and models only edge features without node representations. Its generalization ability to other tasks is weak.
    \item SGCN shows a performance close to SiNE as they said in \cite{derr2018signed}, but it cannot effectively model the signed directed networks because the algorithm did not consider the direction. Its mechanism is the simple average of neighbors' hidden state, which can not describe the influence of different nodes to the target node. 
    \item SiGAT$_{\mathrm{only +/-}}$ shows a performance close to or slightly better than SGCN, indicating that the attention mechanism can better model the problem. Compared with the signed graph convolution network method, attention mechanism considers the influence of different nodes on the target node, which is reasonable.
    \item Our SiGAT model achieves almost the best performance on three datasets. It shows that the relevant theories of sociology have also been successfully modeled by designing different motifs and attention mechanism.
\end{itemize}

\subsection{Parameter Analysis}
\vspace{-20px}
\begin{figure}[H]
\setlength{\leftskip}{-40pt}
   \noindent \makebox[1.26\textwidth][c] {    
    \subfloat[Epoch in Bitcoin-Alpha]{%
      \includegraphics[width=0.32\textwidth]{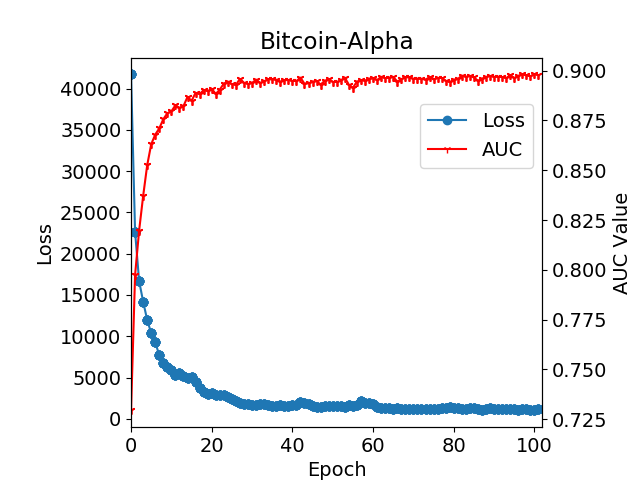}
      \label{fig:1a}
    }
    \subfloat[Epoch in Slashdot]{%
        \includegraphics[width=0.32\textwidth]{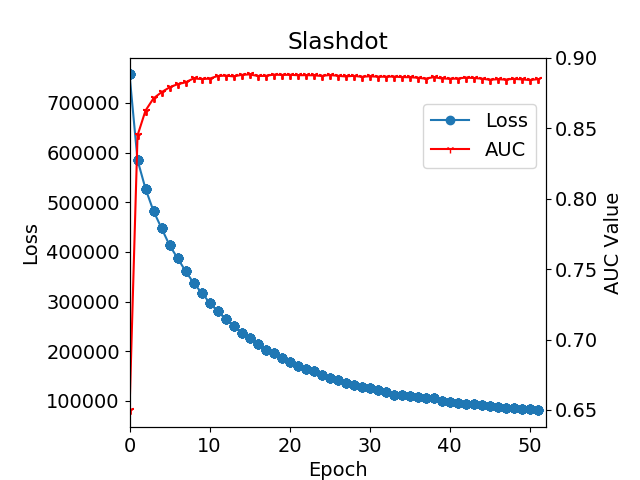}
        \label{fig:1b}
    }
    \subfloat[Epoch in Epinions]{%
        \includegraphics[width=0.32\textwidth]{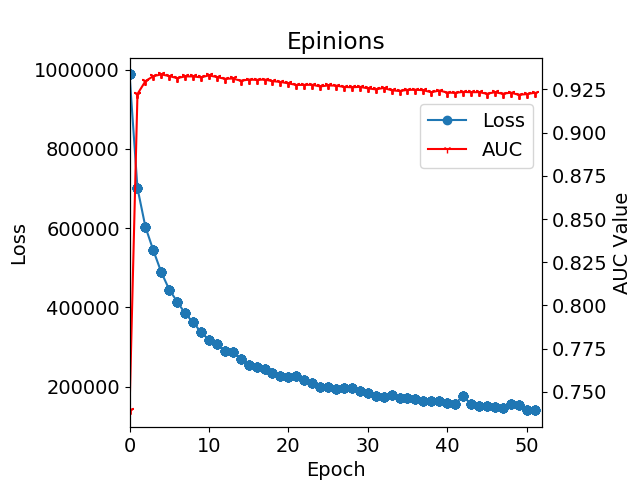}
        \label{fig:1c}
    }
    \subfloat[Dimension]{%
        \includegraphics[width=0.3\textwidth]{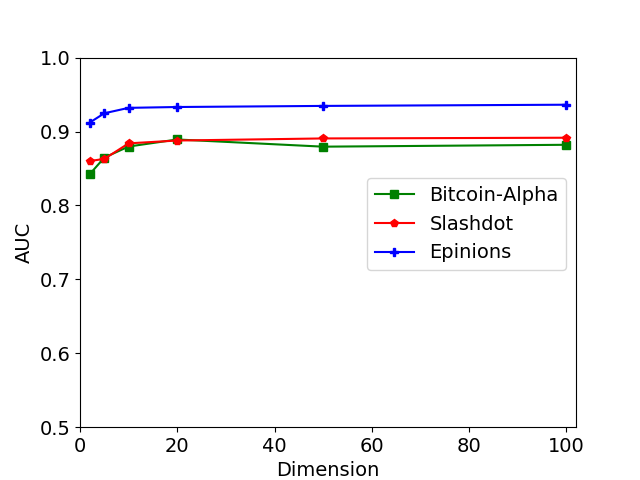}
        \label{fig:1d}
    }
    }
    \caption{Parameter Analysis for the epoch and dimension in three datasets}
    \label{fig:auc_loss}
 \vspace{-10px}
\end{figure}
In this subsection, we analyze the some hyper-parameters. 
We randomly select 80\% training edges and 20\% testing edges.
When analyzing epoch, we set $d=20$, and record the AUC performance of different epoch generated representations and the total loss value during training.
When discussing dimension, we set the corresponding epoch number to 100~(Bitcoin-Alpha), 20~(Slashdot), 10~(Epinions) to discuss the robustness of different dimensions. 

As shown in Fig.\ref{fig:auc_loss}, we can see that the value of loss decreases and the value of AUC increases at the earlier epochs and then gradually converges. In different datasets, the best epochs are different. This can be due to the different scales and network structure. Many nodes are updated multiple times quickly due to the mini-batch algorithm. 
In Fig.\ref{fig:1d}, the performance increases quickly and becomes almost stable with the increasing number of dimension $d$. In conclusion, our model shows relatively robust results on hyper-parameters.

\section{Conclusion}\label{sec:conclusion}
In this paper, we try to introduce GNNs into signed networks. 
GNNs have recently received widespread attention and achieved great improvements in many tasks.
However, most GNNs have been on unsigned networks.
We analyzed some key points in signed network analysis and redesigned the GNNs architectures. 
Based on the basic assumptions of social psychology, we leverage the attention mechanism and different motifs to model the signed network. 
Experimental results on three real-world signed networks show that our proposed SiGAT outperforms different state-of-the-art signed network embedding methods. 
Moreover, we analyze the hyper-parameters and show the robustness of our methods.
In future work, we will further investigate the performance of SiGAT on more signed network tasks, such as node classification and clustering. Further, we will consider more effective methods to model signed networks.

\section*{Acknowledgement}
This work is funded by the National Natural Science Foundation of China under grant numbers 61425016, 61433014, and 91746301. Huawei Shen is also funded by K.C. Wong Education Foundation and the Youth Innovation Promotion Association of the Chinese Academy of Sciences.

\bibliographystyle{splncs04}
\bibliography{refs}

\end{document}